\documentclass[10pt,twocolumn]{article}

\usepackage{multicol}
\usepackage{graphicx}
\usepackage{amsmath}
\usepackage{lipsum}
\usepackage{color}
\usepackage{siunitx}
\usepackage{authblk}
\usepackage{subfigure}
\usepackage[margin=2cm]{geometry}

\title{Drop impact dynamics on slippery liquid-infused porous surfaces: influence of oil thickness}
\author[1,2]{M\'egane Muschi}
\author[2]{Barbara Brudieu}
\author[1,2]{J\'er\'emie Teisseire}
\author[1]{Alban Sauret}
\affil[1]{Surface du Verre et Interfaces, UMR 125 CNRS/Saint-Gobain, 93303 Aubervilliers, France (E-mail: alban.sauret@saint-gobain.com)}
\affil[2]{Saint-Gobain Recherche, 93303 Aubervilliers, France}
\date{ }
\begin{document}

\twocolumn[
    \begin{@twocolumnfalse}
        \maketitle
        \begin{abstract}
            Slippery liquid-infused porous surfaces (SLIPS) are porous nanostructures impregnated with a low surface tension lubricant. They have recently shown great promise in various applications that require non-wettable superhydrophobic surfaces. In this paper, we investigate experimentally the influence of the oil thickness on the wetting properties and drop impact dynamics of new SLIPS. By tuning the thickness of the oil layer deposited through spin-coating, we show that a sufficiently thick layer of oil is necessary to avoid dewetting spots on the porous nanostructure and thus increasing the homogeneity of the liquid distribution. Drop impact on these surfaces is investigated with a particular emphasis on the spreading and rebound dynamics when varying the oil thickness and the Weber number. \\
            \medskip\medskip\medskip
        \end{abstract}
    \end{@twocolumnfalse}
]

\section{Introduction}

In nature, a wide range of biological surfaces exhibit extreme surface characteristics, such as the water repellent properties of some insects and plants \cite{Ref1,Ref2,Ref3}. Various studies have focused on mimicking these natural water repellent surfaces for applications such as biomedical devices, waterproofing clothes, concrete, or glass \cite{Ref4,Ref5,Ref6,Ref7}. For example, inspired by the textured lotus leaf, engineered superhydrophobic (SH) surfaces exhibit remarkable properties such as anti-biofouling, self-cleaning (the so-called lotus effect), anti-icing, and drag reduction due to their surface chemistry or geometry \cite{Ref2,Ref7,Ref8}. Unfortunately, SH surfaces have critical limitations: poor mechanical resilience, low transparency, and weak stability under operating conditions such as repeated drop impacts \cite{Ref9,Ref10}. Indeed, the air pockets contained in the microtexture, which are responsible for SH properties, can easily be filled by liquid under high-pressure conditions, leading to a change in the wetting properties and the loss of the slippery properties. The drop, initially placed on a thin film of air on the surface (\textit{i.e.}, in the Cassie state) becomes impaled into the texturation in a Wenzel state \cite{Ref11}. Therefore, these surfaces are not reliable over a long time scale, especially for outdoor applications such as a car windshield.

To obtain more reliable surfaces, the synthesis of a new kind of surfaces, named Slippery Liquid Infused Porous Surfaces (SLIPS or LIS), in which air pockets are replaced by a low surface tension lubricating film, was recently proposed \cite{Ref12,Ref13}. The liquid, typically oil, is trapped in the pores of the rough surface by capillarity and leads to a smooth and homogeneous surface with a small contact angle hysteresis (typically smaller than a few degrees) and strong slippery properties. In addition to the low contact angle hysteresis, SLIPS exhibit self-cleaning, self-healing, anti-icing properties and are also able to repel various liquids with lower surface tension than water \cite{Ref12,Ref14,Ref15,Ref16,Ref17}. Based on these previous observations, SLIPS appear suitable for a wide variety of commercial and technological applications since drops roll on these surfaces when inclined by a few degrees \cite{gas2017drop}. Additionally, they limit ice formation and have the ability to self-heal due to the oil trapped in the porous structure, which can replace the oil in the surface by capillarity \cite{Ref12,Ref18}. The properties of SLIPS are not expected to change as long as oil is present in the pores and above the top surface, which increases the mechanical stability compared to classical solid SH surfaces.

Only a few studies have considered the behavior of SLIPS when impacted by drops, despite the promising potential of these surfaces. The study of SLIPS is especially important as SH surfaces perform poorly. Previous studies have investigated the effect of oil viscosity on water drop impact dynamic \cite{Ref19,kim2016droplet} or sliding \cite{gas2017drop}. The viscosity of the oil has been shown to slightly affect the maximal spreading radius but has a stronger effect on the drop retraction rate (defined as the retraction speed divided by the maximum radius). To investigate the interactions between the drop and the liquid film during impact, Lee \textit{et al.} varied the Weber number, which compares the inertial effects to the capillary effects \cite{Ref19}. They reported that the splashing threshold, corresponding to destabilization of the outer liquid rim, appears at a larger Weber number when increasing oil viscosity \cite{Ref19,Ref20}.

In this paper, we investigate experimentally the effect of oil thickness on the slippery properties of new liquid-infused surface made with a porous media. To control the porosity and obtain a thick porous layer that increases the durability of the surfaces, we used a sol-gel synthesis and an organic porogen agent coated with a fluorosilane, which is presented in section \ref{Experimental_methods}. The oil thickness is tuned using spin-coating deposition. We characterize the static properties of a drop deposited on a SLIP surface in section 3. We then focus on the drop impact dynamics and discuss the experimental observations in section \ref{Drop_impact}. Our results demonstrate how the oil thickness affects the surface quality. We also show that the spreading phenomena, as well as the rebound dynamic, are weakly influenced by the oil thickness in the range of thickness considered in this study where no large dewetting is observed (between 500 rpm and 2500 rpm).

\section{Experimental methods} \label{Experimental_methods}

\subsection{Surface preparation}

\begin{figure}
    \centering
\includegraphics[width=0.48\textwidth]{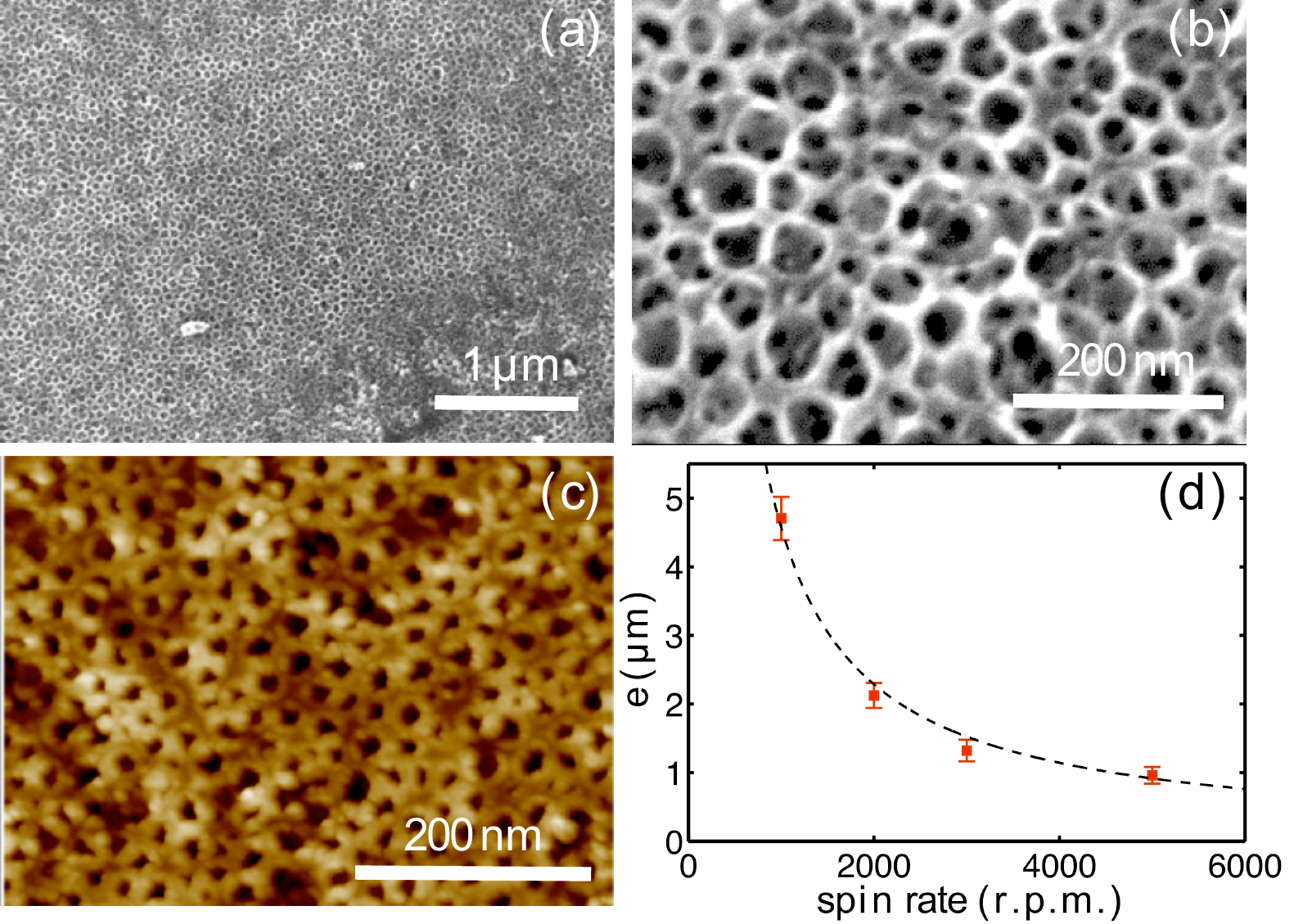}
\caption{(a)-(b) Scanned Electron Microscope (SEM) images of a vertical slice of the nano-porous layer. (c) Atomic-force microscopy (AFM) showing the surface of the porous layer. (d) Oil thickness $e$ when varying the spinning rate (Red squares are data extracted from Ref. 15). The dashed line shows the scaling $e \propto 1/\omega$ where $\omega$ is the spin rate.}\label{figure1}
\end{figure}

The SLIPS are made of a glass substrate on which microporous multilayers silica are deposited. A porogen agent is used to make the porous structure \cite{Ref21}. We synthesize PMMA nanoparticles of 60 nm diameter by radical emulsion polymerization. The layer is made using two silica precursors, glycidoxypropyltriethoxysilane (GLYMO) and tetraethylorthosilicate (TEOS), which are separately hydrolyzed in acidic conditions for 3h30 at room temperature. The condensation reaction is carried out at $60^{\rm o}$C for 1h. After cooling to room temperature, the PMMA suspension is added to reach a porosity of 60\%. The obtained suspension is then filtered through a $0.45\,\mu{\rm m}$ nylon membrane and deposited on a glass substrate by spin coating at 2000 rpm for 60 sec, leading to layer of thickness $0.66 \pm 0.03\,{\rm \mu m}$ (before calcination). To obtain a thick porous layer while avoiding the apparition of cracks in the porous structure, we perform a layer by layer deposition. The glass surface is heated to $100^{\rm o}C$ during 2 min between each layer deposition for pre-condensation. After the deposition of 5 layers, the surface is put in an oven at $100^{\rm o}{\rm C}$ for 1h and then calcinated at $450^{\rm o}{\rm C}$ for another hour to degrade the organic part of the coating and obtain the desired porosity. Surfaces are then treated with UVO$_3$ for 1h. The resulting porous silica layer is hydrophilic and has thus no affinity to the hydrophobic oil. Therefore, before oil impregnation, the surface must be made hydrophobic by grafting a fluorosilane molecule on the surface using vapor deposition. More specifically, the fluorosilanization was carried out by adding 20 $\mu{\rm L}$ of (1H,1H,2H,2H-perfluorododecyltrichlorosilane) in vapor phase under nitrogen atmosphere in a desiccator prior purged with the surfaces. The grafting is made by vapor deposition of the fluorosilane on the surfaces for 4h under static vacuum. Because the 20 $\mu{\rm L}$ are added in the vapor phase, all the (1H,1H,2H,2H-perfluorododecyltrichlorosilane) does not contribute to the final surface but the resulting properties of the surfaces are reproducible.

We then use a spin-coating method for the oil deposition to control the thickness of the layer through the rotation rate. Krytox 100 oil (viscosity of 10 cSt) is used based on the literature and referred to as lubricant \cite{Ref12,Ref15,Ref16,Ref22,Ref23}. The lubricant is filtered with a 1 $\mu$m nylon membrane and deposited in two steps. First, the oil is infused in the porous layer by depositing an excess of oil and spin-coating the sample at 1000 rpm for 60 sec. Then, the oil thickness $e$ is varied by depositing a second-time oil and spin-coating at a rotation rate in the range 200 to 5000 rpm. Hereafter, we refer to the rotation rate of this second stage only as we observed that the first deposition has no significant impact on the behavior of the SLIP surface.

\subsection{Characterization of the surfaces}

The resulting samples, made of five nanoporous layers, are characterized by scanning electron microscopy (SEM) and atomic force microscopy (AFM) as shown in Fig. \ref{figure1}(a)-(b). The total thickness of the porous structure is $2.1\,\mu {\rm m}$ and no demarcation between the deposited layers is observed, revealing a homogeneous porous layer. To our knowledge, the reported SLIPS usually have a smaller thickness. However, we believe that by successfully obtaining a thick layer, the durability of such surfaces will be improved. The cross-section of the porous layer observed by SEM shows that the pores are interconnected [Fig. \ref{figure1}(a)-(c)]. This interconnection allows for the oil to completely infuse the porous surface, thus increasing the volume of the reservoir of oil and improving the durability of the surface properties. Analysis by AFM reveals a surface porosity of 38\%, a value that is smaller than the theoretical porosity in volume (60\%) because of the low affinity of PMMA with the surface but sufficient to allow the oil impregnation. Here, the 38\% porosity found by AFM is a surface porosity but the experimental volume porosity is the same as the theoretical one, \textit{i.e.}, 60\% as shown in a previous study \cite{Ref21} We also performed ellipsometry measurements, which confirmed that the volume porosity is approximately 60\% in agreement with this study.

In the literature, most deposition methods of the oil consist of dipping the surface into a reservoir of lubricant or by simply spreading the oil with a pipette \cite{Ref14,Ref15,Ref24}. Determining the oil thickness remains challenging. However, a few studies have considered the oil deposition by spin-coating, which appears to be more controlled. They also estimated the thickness values by weighing the surfaces before and after oil impregnation for the oil used here, Krytox 100 \cite{Ref15,Ref24}. From these measurements, we estimate that at low spin rate, as used in this study (500 rpm), the oil thickness is about $8-9\,\mu{\rm m}$ whereas at larger spin rate the oil thickness is lower, around $1\,\mu{\rm m}$ as reported in Fig. \ref{figure1}(d). We emphasize that we here consider the common approach, which is to estimate the thickness of the oil layer on top of the porous substrate and do not consider the oil trapped in the pore that only constitute a reservoir of oil for self-healing purposes. We shall see later that decreasing the oil thickness leads eventually to more dewetting spots on the SLIPS.

\section{Static wetting of the SLIPS}

\begin{figure}
    \centering
\includegraphics[width=0.4\textwidth]{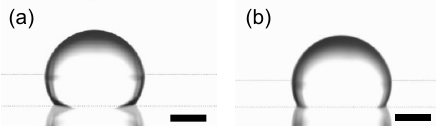}
\caption{Contact angle of a drop of water on the porous layers (a) functionalized with a fluorosilane and (b) functionalized with a fluorosilane and impregnated by fluorinated oil deposited at 2500 rpm. Scale bars are 1 mm.}\label{figure2}
\end{figure}

The contact angle measurements were performed with a PGX contact angle measurement meter using a drop volume of $5.5 \mu{\rm L}$. To ensure that the fluorosilane is grafted to the porous layer and thus that the surface is hydrophobic, the advancing and redecing contact angles are measured before and after oil impregnation. We obtain for the non-infused surface an advancing angle of $\theta_a=140^{\rm o}$ and a receding angle $\theta_r=95^{\rm o}$ leading to a large contact angle hysteresis $\Delta \theta=\theta_a-\theta_r=45^{\rm o}$. A drop of water deposited on such surface is shown in Fig. \ref{figure2}(a), and the obtained values confirm the fluorosilanisation of the porous nanostructure. After oil impregnation by the Krytox 100, the advancing angle becomes $\theta_a=119^{\rm o}$ and the receding contact angle is $\theta_r=115^{\rm o}$, which leads to a small contact angle hysteresis, typically around $\Delta \theta=4^{\rm o}$ for all oil thickness considered up to 2500 r.p.m. [Fig. \ref{figure2}(b)]. We also observe than the same static contact angle is obtained by spreading a large excess of oil on a synthesized SLIPS with a pipette followed by the removal of the excess by tilting the surface for 30 minutes. Therefore, it appears that the thickness of the oil does not significantly affect the static contact angle in the range considered here as observed in previous studies \cite{Ref14,Ref15,Ref23}.

\begin{figure}
    \centering
\includegraphics[width=0.48\textwidth]{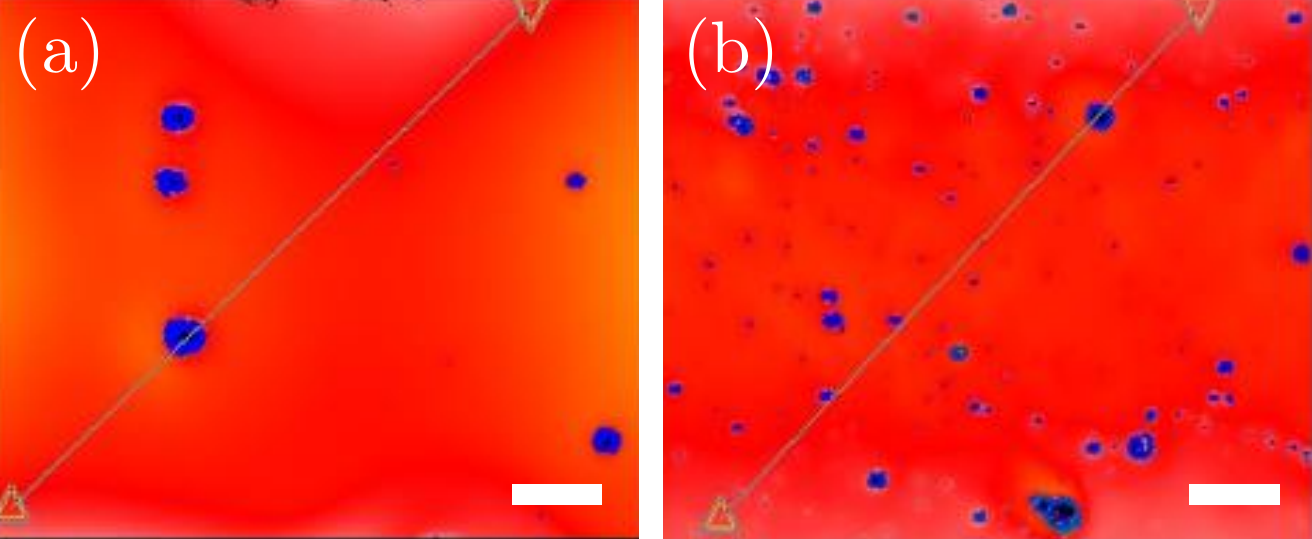}
\caption{Interferometer microscope images of a SLIPS with oil deposited (a) at 500 rpm and (b) at 2500 rpm. The color indicate the relative liquid thickness, the blue color being dewetting spots. Scale bars are $1\,{\rm mm}$.}\label{figure3}
\end{figure}

Two impregnated surfaces of different oil thickness (spin rate during oil deposition of 500 and 2500 rpm) are characterized using an imaging interferometric microscope and are reported in Fig. \ref{figure3}(a)-(b). The oil thickness is qualitatively visible in this figure: the red corresponds to the region where the liquid height is homogeneous and the blue corresponds to region having a smaller height and thus a depletion of oil. The sample with oil deposited at 500 rpm shows a relatively homogeneous surface with few dewetting spots (blue spots). The sample for which oil was deposited at 2500 rpm shows many more dewetting spots, revealing a lower surface quality. These results suggest that thinner oil layers exhibit more defects visible on the surface. However, these dewetting spots do not affect the static contact angle because of their relatively low size, typically few tens to hundred micrometers, compared to the drop diameter (few millimeters).

To characterize the slippery properties of these new surfaces, and investigate the role of the oil thickness on the performance, we also performed sliding angle experiments. The surfaces are characterized as slippery if their contact angle hysteresis (\textit{i.e}, sliding angle) is typically below $5^{\rm o}$. Therefore, we placed $5.5 \mu{\rm L}$ water drops on surfaces on which oil was deposited at rotation rates ranging from 200 to 5000 rpm (200, 500, 1000, 1500, 2000, 2500 or 5000 rpm). The surfaces are initially inclined at $5^{\rm o}$ and we observe the behavior of the drop once deposited on the surface. If the drop remains attached to the surface, we incline the surface further until we reach the smallest angle at which the drop slides. We observed that only the 5000 rpm sample was not slippery with a sliding angle value of 28$^{\rm o}$. For this reason, we focus in the article on surfaces that remain slippery and considered the smaller and larger deposition rate available, \textit{i.e.}, 500 and 2500 rpm. At the largest deposition rate considered, 5000 rpm, the centrifugation pressure, $\rho\,\omega^2\,D^2/8$ ($\rho$: density of the oil, $\omega$: rotation rate, $D$: diameter of the sample surface) is larger than the Laplace pressure $\gamma/2\,R$ ($\gamma$: interfacial tension of the oil, $R$: radius of the pores), leading to the removal of the oil from the surface explaining that for this deposition rate the surface was found to be not slippery. However, although the oil thickness influences the surface homogeneity for the other used speed rate deposition, it does not strongly affect the static contact angle nor the sliding angle.

\section{Results and discussions}  \label{Drop_impact}

\subsection{Phenomenology}

\begin{figure}
    \centering
\includegraphics[width=0.48\textwidth]{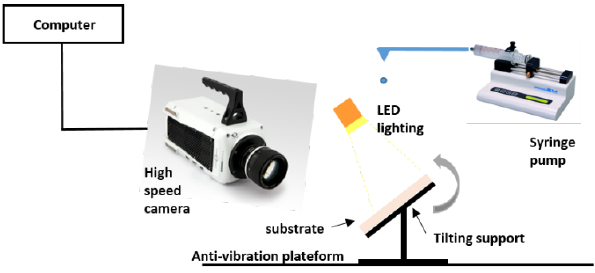}
\caption{Schematic of the experimental set up for drop impact.}\label{figure6}
\end{figure}

To study drop impacts on SLIPS, water drops are generated using a syringe connected to a syringe pump allowing control of the volume of the drops. Here, we used water drops of radius $R_0=1.6\,{\rm mm}$. The drop impact dynamics are observed using a high speed camera (Phantom v611) with a macro lens (Nikon 105 mm), at 10,000 frames per second, allowing capture of the fast dynamic of the drop at the impact and its evolution. The SLIPS are placed on a tilting sample support to study the impacts on flat surfaces. LED panels are placed behind and under the samples to ensure a sufficient lighting of the experiments.

\begin{figure*}
    \centering
\includegraphics[width=0.9\textwidth]{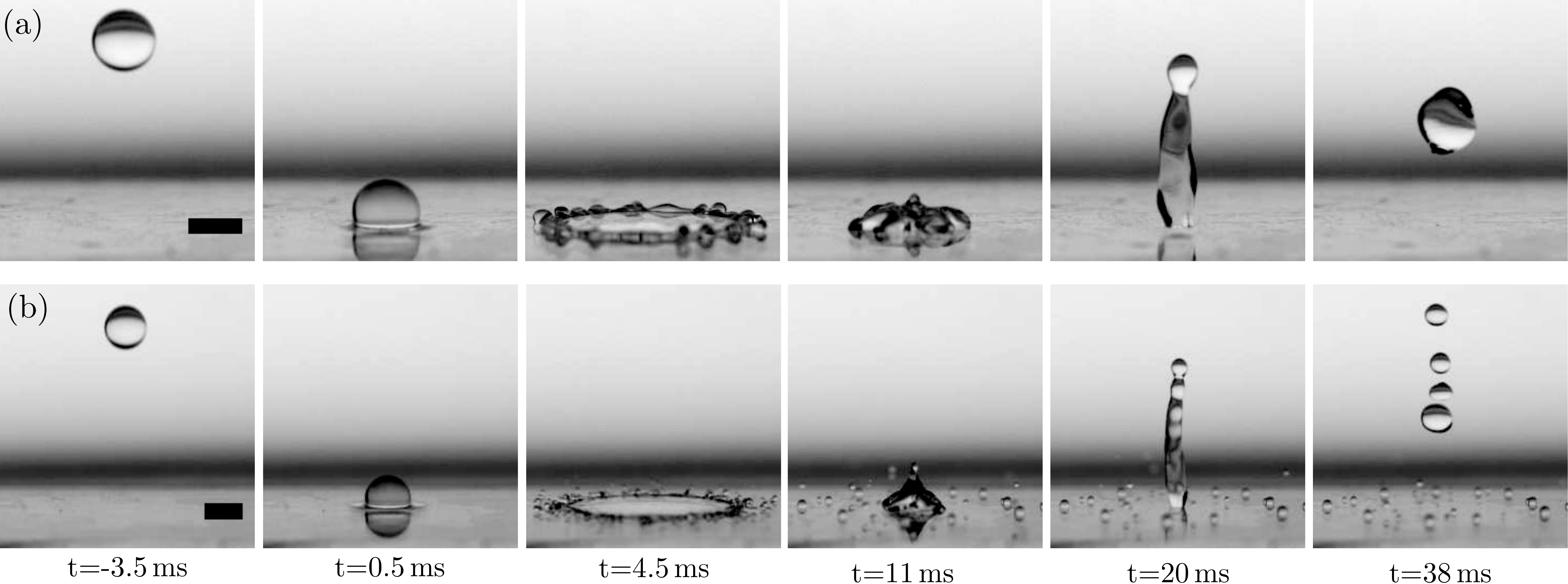}
\caption{Impact and rebound of a water drop of radius $R_0=1.1\,{\rm mm}$ on a SLIPS with oil deposited at 500 rpm for an impact velocity of (a) $U_0=2.2\,{\rm m.s^{-1}}$ and (b) $U_0=4.7\,{\rm m.s^{-1}}$, corresponding to Weber numbers of $We=66$ and $We=422$, respectively. The time scale is the same on both figures. Scale bars are $2\,{\rm mm}$.}\label{figuredyna}
\end{figure*}

The drop impact is characterized by the Weber number, which represents the ratio between the inertial and interfacial tension effects and is defined as $We = \rho\,{U_0}^2\,R_0/\gamma$, where $\rho$ is the water density ($1000\,{\rm kg.m^{-3}}$), $U_0$, $R_0$ are the drop speed and radius, respectively and $\gamma$ is the water/air surface tension ($72\,{\rm mN.m^{-1}}$). To study the impact dynamics as a function of the Weber number, the impact velocity is varied by increasing the drop release height $h$, which leads to a range of Weber number from 66 to 950. To observe the maximal spreading radius, the camera is placed above the sample. For measurements, the external rim is not considered in order to avoid imprecisions between non-splashing drops and splashing drops. For the characterization of rebound and contact time, the camera was placed in the sample axe as shown on the schematic in Fig. \ref{figure6}.

Two characteristic experiments of drop impact at different velocities but on the same SLIPS on which the oil was deposited at 500 rpm, are shown in Fig. \ref{figuredyna}(a)-(b). Both situations look qualitatively similar. Indeed, we observe that, during the first phase of the drop impact, the drop spreads over the surface and forms a liquid film of thickness equal to a few hundred micrometers (between $t=0\,{\rm ms}$ and $t \simeq 4.5 \,{\rm ms}$). The drop reaches its maximum radius of spreading at $t \simeq 4.5\,{\rm ms}$ in both situations. At this stage, we should emphasize that the film thickness is not uniform as a liquid rim forms at the edge and the layer is thinner at the center. After reaching its maximum radius, the liquid film shrinks in size between $t = 5 \,{\rm ms}$ and $t = 11 \,{\rm ms}$, which generates a fluid flow toward the center, leading to a vertical motion and the bouncing of the drop off the surface. Note that for both impact speeds, the contact time of the drop with the surface remains roughly constant and equal to approximatively $ 20 \, {\rm ms}$. A larger impact velocity, shown in Fig. \ref{figuredyna}(b), the bouncing dynamics remain similar even if the drop flattens more during spreading, which leads to a larger maximum radius. This can also lead to splashing at sufficiently large Weber numbers. However, the contact time is almost identical, \textit{i.e.}, independent of the impact velocity.

\subsection{Maximal spreading radius}

\begin{figure}[h!]
    \centering
\includegraphics[width=0.45\textwidth]{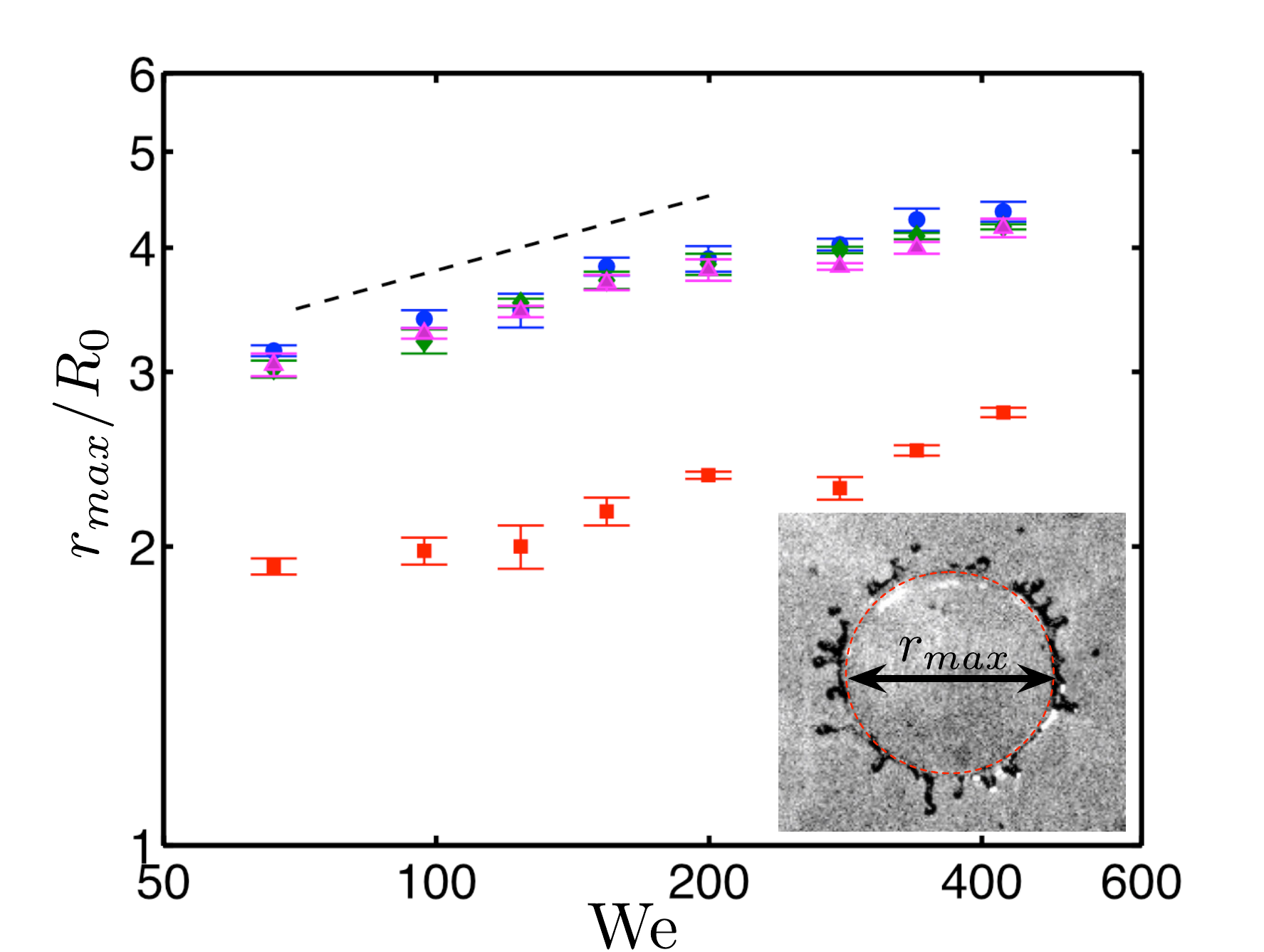}
\caption{Maximal spreading ratio $r_{max}/R_0$, measured inside the crown (see inset), as a function of the Weber number $We$ for SLIPS within which oil is deposited at 500 rpm (blue circles), 1000 rpm (green diamonds) and 2500 rpm (purple triangles). The red square are the results for the porous surface without oil. The dash-dotted line has a slope 1/4.
}\label{figure7}
\end{figure}

When the drop impacts the surface, it accumulates kinetic energy during the fall, leading to the spreading of the droplet onto the surface. The drop spreading is determined by the balance be- tween capillarity and inertia \cite{Ref6,Ref25}. When the capillary energy and kinetic energy become equal, the drop is at its maximum spreading diameter. The measurements of $r_{max}$ obtained experimentally on three slippery surfaces with different oil speed deposition are shown in Fig. \ref{figure7}. We also report the value for the porous layer without oil. We observe that $r_{max}$ increases with the Weber number, and thus with the impact velocity. The value of $r_{max}$ at a given Weber number is approximately the same for all samples covered with oil, suggesting that the oil thickness does not have any effect on $r_{max}$ in the range considered in our study. We should emphasize here that we consider the maximum radius measured inside the crown, whereas in the next section the dynamics of the outer crown are considered.

This observation can be attributed to the fact that viscous dissipation inside the drop is larger than viscous dissipation inside the thin oil layer. Indeed, following the arguments developed by Lee \textit{et al.}, who considered the influence of the viscosity of the oil on the maximum spreading radius, we can compare the viscous dissipation inside the water drop and inside the oil layer in the present situation \cite{Ref19}. When the drop impacts the SLIPS, the viscous force in the oil layer is $ \eta_o\,({\partial^2 U_{oil}})/({\partial y^2}) \sim \eta_o\,{U_{oil}}/{e^2}$, where $\eta_0$ is the dynamic viscosity of the oil, and $U_{oil}$ and $e$ are the characteristic viscosity and the thickness of the oil layer, respectively. The viscous force is balanced by the dynamic pressure induced by the impact of the drop and equal to $\rho_w\,{U^0}^2/R_0$. We therefore obtained an estimate of the characteristic velocity in the oil layer produced at the drop impact: $U_{oil} \sim ({\rho_W\,e^2\,{U_0}^2})/({\eta_o\,R_0})$. We can then estimate the viscous dissipation in both the drop and the oil layer during the spreading phase. The viscous dissipation in the water drop is of order $\eta_w\,(U-U_{oil})^2\,{r_{max}}^2/h$, where $U$ is the mean drop spreading velocity and $h$ is its thickness (typically a few hundred of micrometers). In the oil layer, the viscous dissipation is of order $\eta_o\,{U_{oil}}^2\,{r_{max}}^2/e$. We can then express the ratio of the viscous dissipation inside the thin oil layer to the viscous dissipation inside the drop during its spreading on the surface, which is equal to
\begin{equation} \label{relation_ratio}
\frac{\eta_o\,{U_{oil}}^2\,{r_{max}}^2/e}{\eta_w\,(U-U_{oil})^2\,{r_{max}}^2/h} \sim \frac{U_{oil}}{U-U_{oil}} \sim \frac{\eta_w}{\eta_o} \,\frac{e}{h},
\end{equation}
where we have used the continuity of stress at the water-oil interface through the relation $\eta_w\,(U-U_{oil)})/h \sim \eta_o\,U_{oil}/e$. Estimating a range of values for the different parameters, we have $\eta_w/\eta_0\sim 0.1$, $e \leq 10\,\mu{\rm m}$ and $h \sim 100\,{\rm \mu m}$. Relation (\ref{relation_ratio}) shows that the viscous dissipation in the oil layer during the spreading stage is negligible compared to the viscous dissipation in the water drop. Therefore, the oil thickness $e$ has no quantitative effect on the maximal spreading radius of the drop $r_{max}$ as observed in Fig. \ref{figure7}.

Our experimental results also show that the maximal spreading radius follows a $We^{1/4}$ scaling-law as observed for some SH and infused surfaces \cite{Ref19,Ref26}. This law is still subject to debate but has proven to agree with most experimental observations \cite{Ref19,Ref27,Ref28}. The dependence of the maximum radius $r_{max}$ with the Weber number can be explained through the equilibrium between the kinetic energy at the impact and the surface tension of the drop. The $We^{1/4}$ scaling law is induced by the sudden drop deceleration at impact. More specifically, when the deformation of the drop is maximal and becomes nearly flat, the gravity force overcomes the surface tension force. At the impact on the SLIP surface, the drop of water of radius $R_0$ decelerates from $U_0$ to $0$ in a time scale equal to $\tau_{dec}=2\,R_0/U_0$. Therefore, the sudden deceleration experienced by the drop at the impact scales as ${U_0}^2/R_0$. This deceleration leads to an apparent gravity field $g*$, much larger than $g$, and of order $g* \sim {U_0}^2/R_0$. Using this expression of the acceleration in the capillary length, the thickness$e$ of the drop when it reaches its maximum radius $r_{max}$ is $e=\sqrt{\gamma/(\rho\,g^*)}$. Then, a volume conservation principle writes $\pi\,{r_{max}^2}\,e=4\,\pi\,{R_0}^3/3$. Using the expression of $e$ and $g^*$ and the Weber number $We$, we obtain the scaling law $r_{max} \propto R_0\,We^{1/4}$ \cite{Ref26}. We also notice that a small stagnation of $r_{max}/R_0$ is observed between $We = 200$ and $We = 279$. This stagnation can be attributed to the apparition of prompt splashing which leads to the ejection of microdroplets from the rim of the sheet and thus affects the maximal spreading radius. 

Finally, we can observe that, in absence of oil, the value of $r_{max}/R_0$ is smaller than for all oil thicknesses considered. Indeed, in this situation, the drop impacts a porous media, which changes the dynamics because the thin film of air forming under the drop at the impact can be affected by the empty pores. However, if we consider the maximum radius of the crown of the impacting drop as we shall see in the next section, the difference is much smaller. We believe that this observation is a result of the type of substrate (porous versus thin liquid layer) that leads to a different dynamic of the crown at the impact.

\subsection{Spreading and retraction dynamics}

We observe that the spreading and retraction dynamics are not symmetric. Indeed, the drops spread much faster ($\sim\,4.5$ ms) on the surface than they retract ($\sim\,16$ ms). Following Clanet \textit{et al.} \cite{Ref26}, we define the characteristic time scale $\tau=\sqrt{\rho \, {R_0}^3 / \sigma}$, equals to $\tau \simeq 7.5\,{\rm ms}$. For both cases shown in Fig. \ref{figuredyna}, the drops reach their maximum radius at the same time. Therefore, at larger impact speed, the drops spread faster. The same situation is visible for the shrinking from this maximum radius; the drop retracts over a greater distance when they impact at large speed, but their shrinking speed is faster than the speed observed at low impact velocity.

\begin{figure}
    \centering
\includegraphics[width=0.48\textwidth]{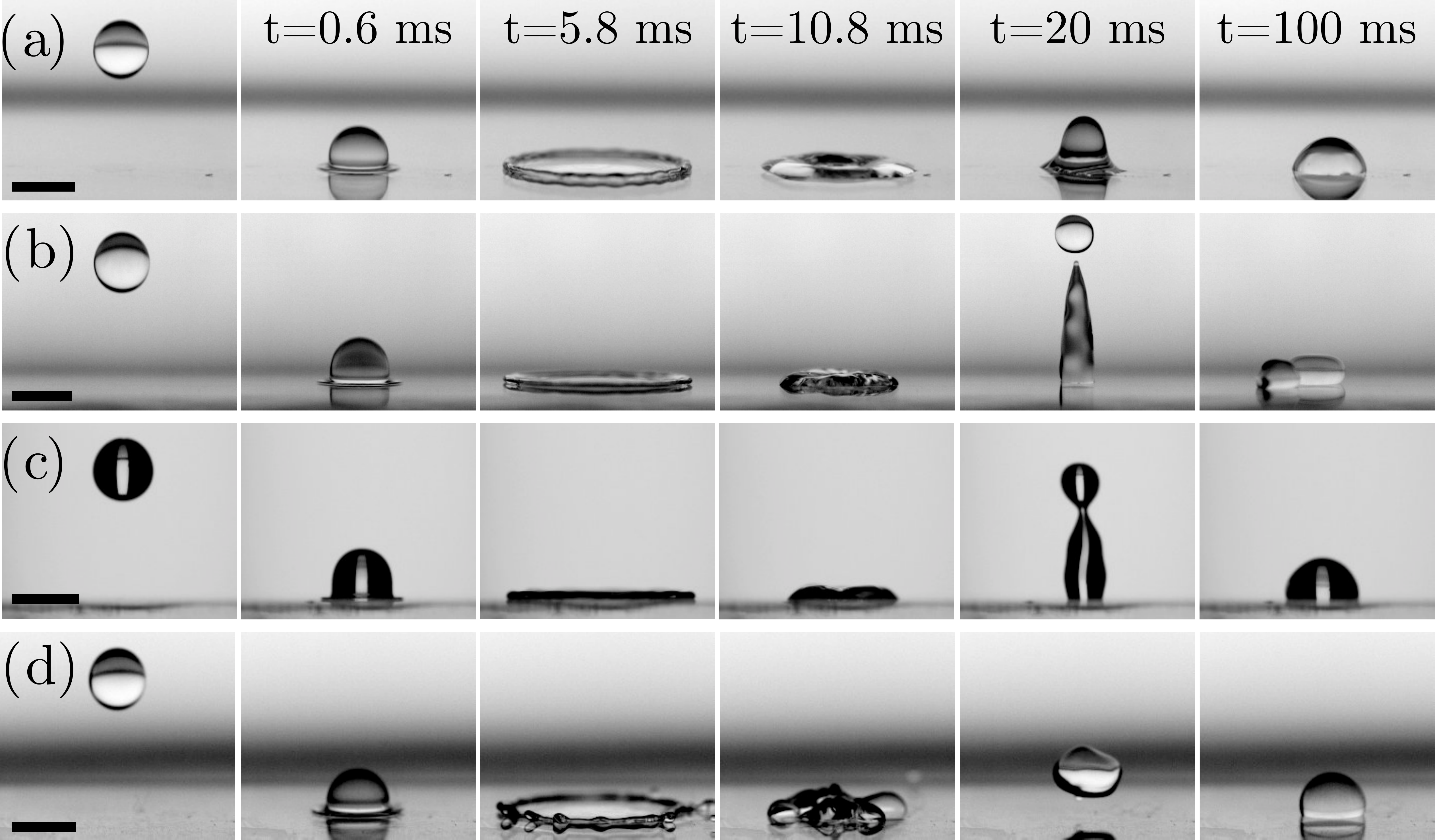}
\caption{Time evolution of a water drop impacting different test surfaces at a Weber number of We = 66 : (a) hydrophobic coating (fluorinated silane grafted to glass), (b) porous surface without oil and SLIPS with oil deposited at (c) 5000 rpm and (d) 2500 rpm. Scale bars are $3\,{\rm mm}$.
}\label{figureIllustration}
\end{figure}

We also performed series of experiments investigating the spreading and the retraction dynamics of the water drops on different surfaces. Examples of these experiments are reported in Fig. \ref{figureIllustration}. We first consider an industrial water-repellent coating, which consists of fluorinated silane grafted to glass, and turns glass permanently hydrophobic. The contact angle on this surface is $110^{\rm o} \pm 5^{\rm o}$ and the contact angle hysteresis is about $30^{\rm o} \pm 5^{\rm o}$. Therefore, the contact angle is similar to the contact angle measured on our SLIPS, whereas the contact angle hysteresis is much larger. We observe in Fig. \ref{figureIllustration}(a), that a drop impacting this coating does not bounce in contrast to what is observed for the SLIPS presented in Fig. \ref{figuredyna}. The absence of bouncing is also observed on the porous substrate (without oil) as shown in Fig. \ref{figureIllustration}(b). We also compared two SLIPS with different thicknesses of oil, deposited at 5000 rpm and 2500 rpm. We observe that when the thickness of the oil layer becomes too small, no bouncing is observed (Fig. \ref{figureIllustration}(c)). Thus, we will only consider oil layers deposited between 500 rpm and 2500 rpm to ensure the bouncing of the drop and compare the influence of the oil thickness.

\begin{figure}[h!]
    \centering
\includegraphics[width=0.4\textwidth]{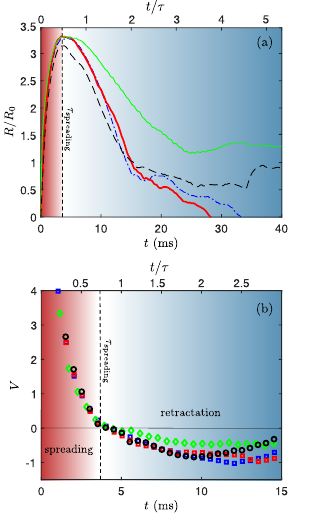}
\caption{(a) Time evolution of the diameter of the impacting water drop normalized by the initial drop diameter at a Weber number of We = 66 on different substrates: fluorinated silane grafted to glass (green line), porous substrate without oil (dotted black), SLIPS at 500 rpm (dash-dotted blue) and 2500 rpm (thick red line). (b) Time evolution of the spreading and retraction velocities of droplet impacting on the same surfaces than in (a) at a Weber number of We = 66. $\tau$ is the characteristic time equals to 7.5 ms}\label{Position_Speed}
\end{figure}

More quantitatively, we investigated the time evolution of the drop radius when it impacts and spreads on the surfaces (Fig. \ref{Position_Speed}(a)). These data are extracted from the side view and therefore includes the external rim. We do not observe a strong difference between the surface coating with the fluorinated silane grafted to glass, the porous layer, and the SLIPS with oil deposited at 2500 rpm or 500 rpm. The main difference appears during the receding phase where the large contact angle hysteresis on the surface coated with fluorinated silane grafted to glass leads to a slower velocity, as emphasized in Fig. \ref{Position_Speed}(b) where we report the spreading and receding velocities calculated from the droplet diameter data. We observe that, whereas the spreading and receding velocity seems comparable on the porous substrate and on the SLIPS, the maximum diameter reached by the drop is smaller, and the drop does not bounce on the surface. Quantitatively, we did not observe a significant difference between a coating at 2500 rpm and 500 rpm. We will investigate this similarity in the next section.

\subsection{Contact time between the drop and the surface}

\begin{figure}[h!]
    \centering
\includegraphics[width=0.425\textwidth]{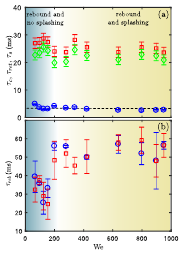}
\caption{(a) Contact time (red squares), retraction time (green diamonds) and spreading time (blue circles) as a function of the Weber number for a slippery liquid infused surface within which oil was deposited at 2500 rpm. (b) Rebound time for varying Weber number for two SLIPS with oil deposited at 500 rpm (red squares) and 2500 rpm (blue circles). The blue and yellow regions indicates when the drop impacts without splashing and with splashins, respectively, and then rebound.}\label{figure8}
\end{figure}

The contact time of the drop with the surface, \textit{i.e.}, the sum of the spreading and the retraction times, is relevant to characterize the drop impact dynamics, considering that contact time depends on the inertia and capillarity forces of the drop, which are interactions with the surface and dissipation. Different studies have focused on minimizing this contact time for applications such as anti-icing coating on SH surfaces \cite{Ref29,Ref30}. To consider this parameter of SLIPS, we measured the contact time on a sample on which oil was deposited at 2500 rpm and reported the values in Fig. \ref{figure8}(a). In the range of Weber number considered, for a given radius of drop, we found that the contact time is independent of the impact velocity in agreement with past studies performed on superhydrophobic surfaces \cite{Ref31}.

The retraction phase is followed by the drop bouncing. We investigated the rebound time, \textit{i.e.}, the time during which the drop bounces off the surface, as a function of the Weber number on two liquid-infused surfaces coated at 500 rpm and 2500 rpm in Fig. \ref{figure8}(b). In the range of Weber number considered, our results reveal that the drops bounce off the SLIPS even at relatively low We numbers. These results are specific to SLIPS since it is known that water drops do not bounce on surfaces with receding contact angles higher than $100^{\rm o}$ as shown by Antonini \textit{et al.} \cite{antonini2013drop,Ref32,malavasi2016knowledge}. Our measurements show that the receding angle for the non-infused surface, $\theta_r=95^{\rm o}$, does not satisfy this criterion and therefore it explains why the drop does not bounce on such surface. However, for the SLIP surface infused with Kritox 100 the receding angle is larger and equal to $\theta_r=115^{\rm o}$, which explains the bouncing observed in our experiments. In the litterature, this bouncing dynamic was attributed to the presence of the oil film, which leads to the reduction of the energy dissipation during the contact phase compared to a classical solid hydrophobic surface. First, the contact angle hysteresis is very low, meaning that the dissipation caused by the drop deformation during spreading and receding is weak. In addition, the frictional forces are weaker on SLIPS than on solid surfaces. Therefore, after receding, the drop still has enough energy to bounce off, which is not the case on fluorinated porous surface without oil and surfaces obtained at 5000 rpm. Here, two trends can be observed: (i) below $We \sim 230$, the rebound time fluctuates about $30$ ms for both samples and (ii) above $We = 200$, this average value increases. At $We \sim 230$, the splash phenomenon appears, leading to the ejection of microdroplets and thus to the loss of energy.

\section{Conclusion}

In this paper, we investigated the spreading and the retraction dynamics following a drop impact on slippery liquid infused porous surfaces (SLIPS) using high-speed imaging. Experimentally, we have synthesized a new SLIP surface with a thick porous structure using a sol-gel synthesis method. The oil thickness on the surface was then varied by tuning the spin rate during the deposition of the oil. 

Interferometer microscope images have shown that the oil thickness on the surface has a strong impact on the surface homogeneity. We observed that the contact angle of a drop of water deposited on the SLIPS surface does not depend on the oil thickness in the range of values considered here when the liquid at the surface is mostly homogeneous. In addition, the slippery dynamical properties of our SLIPS were also found not to depend on the oil thickness. 

Drop impact was studied in a large range of Weber numbers to investigate the influence of the oil thickness on the impact dynamics on a flat surface. We observed a drop bounce on the SLIPS surface, whereas no bouncing is observed on surface showing a similar contact angle but a larger contact angle hysteresis. We also found that neither the drop spreading nor the bouncing dynamics were strongly affected by the oil thickness provided that the oil layer remains mostly homogeneous. The maximum spreading diameter exhibited a $We^{1/4}$ scaling law on the SLIPS surface, in the range of oil thickness considered. Interestingly, the retraction rate of the droplet on the SLIPS surface was found to remain mainly constant for different oil thickness. 

Although the oil thickness does not affect the static nor the dynamic properties studied here, it is likely that the durability of the surfaces should depend on this parameter. Therefore, attention should be given to SLIPS design, especially for outdoor applications on a tilted surface. In summary, the oil drainage as well as the presence of dewetting spots could eventually lead to the loss of the slippery properties, and this effect is more important for an initial small thickness of the oil layer.

\section*{Acknowledgments}

The authors acknowledge the technical support of Emmanuel Garre and measurements of the contact angles performed by Vincent Perrot. We also thank an anonymous referee for helpful comments that improve the clarity of the manuscript.

\bibliography{Megane_Ref} 

\begin{thebibliography}{10}

\bibitem{Ref1}
X.~Gao and L.~Jiang, ``Biophysics: Water-repellent legs of water striders,''
  {\em Nature}, vol.~432, no.~7013, pp.~36--36, 2004.
\newblock 10.1038/432036a.

\bibitem{Ref2}
R.~F\"urstner, W.~Barthlott, C.~Neinhuis, and P.~Walzel, ``Wetting and
  self-cleaning properties of artificial superhydrophobic surfaces,'' {\em
  Langmuir}, vol.~21, no.~3, pp.~956--961, 2005.

\bibitem{Ref3}
B.~Bhushan and Y.~C. Jung, ``Natural and biomimetic artificial surfaces for
  superhydrophobicity, self-cleaning, low adhesion, and drag reduction,'' {\em
  Progress in Materials Science}, vol.~56, no.~1, pp.~1--108, 2011.

\bibitem{Ref4}
B.~J. Privett, J.~Youn, S.~A. Hong, J.~Lee, J.~Han, J.~H. Shin, and M.~H.
  Schoenfisch, ``Antibacterial fluorinated silica colloid superhydrophobic
  surfaces,'' {\em Langmuir}, vol.~27, no.~15, pp.~9597--9601, 2011.

\bibitem{Ref5}
K.~Ramaratnam, V.~Tsyalkovsky, V.~Klep, and I.~Luzinov, ``Ultrahydrophobic
  textile surface via decorating fibers with monolayer of reactive
  nanoparticles and non-fluorinated polymer,'' {\em Chemical Communications},
  no.~43, pp.~4510--4512, 2007.

\bibitem{Ref6}
M.~Callies and D.~Quere, ``On water repellency,'' {\em Soft Matter}, vol.~1,
  no.~1, pp.~55--61, 2005.

\bibitem{Ref7}
S.~Yu, Z.~Guo, and W.~Liu, ``Biomimetic transparent and superhydrophobic
  coatings: from nature and beyond nature,'' {\em Chemical Communications},
  vol.~51, no.~10, pp.~1775--1794, 2015.

\bibitem{Ref8}
X.~Zhang, F.~Shi, J.~Niu, Y.~Jiang, and Z.~Wang, ``Superhydrophobic surfaces:
  from structural control to functional application,'' {\em Journal of
  Materials Chemistry}, vol.~18, no.~6, pp.~621--633, 2008.

\bibitem{Ref9}
D.~Qu\'er\'e, ``Wetting and roughness,'' {\em Annual Review of Materials
  Research}, vol.~38, no.~1, pp.~71--99, 2008.

\bibitem{Ref10}
Z.~Guo, W.~Liu, and B.-L. Su, ``Superhydrophobic surfaces: From natural to
  biomimetic to functional,'' {\em Journal of Colloid and Interface Science},
  vol.~353, no.~2, pp.~335--355, 2011.

\bibitem{Ref11}
A.~B.~D. Cassie and S.~Baxter, ``Wettability of porous surfaces,'' {\em
  Transactions of the Faraday Society}, vol.~40, no.~0, pp.~546--551, 1944.

\bibitem{Ref12}
T.-S. Wong, S.~H. Kang, S.~K.~Y. Tang, E.~J. Smythe, B.~D. Hatton, A.~Grinthal,
  and J.~Aizenberg, ``Bioinspired self-repairing slippery surfaces with
  pressure-stable omniphobicity,'' {\em Nature}, vol.~477, no.~7365,
  pp.~443--447, 2011.
\newblock 10.1038/nature10447.

\bibitem{Ref13}
A.~Lafuma and D.~Qu\'er\'e, ``Slippery pre-suffused surfaces,'' {\em EPL},
  vol.~96, no.~5, p.~56001, 2011.

\bibitem{Ref14}
J.~D. Smith, R.~Dhiman, S.~Anand, E.~Reza-Garduno, R.~E. Cohen, G.~H. McKinley,
  and K.~K. Varanasi, ``Droplet mobility on lubricant-impregnated surfaces,''
  {\em Soft Matter}, vol.~9, no.~6, pp.~1772--1780, 2013.

\bibitem{Ref15}
N.~Vogel, R.~A. Belisle, B.~Hatton, T.-S. Wong, and J.~Aizenberg,
  ``Transparency and damage tolerance of patternable omniphobic lubricated
  surfaces based on inverse colloidal monolayers,'' {\em Nature
  communications}, vol.~4, p.~2167, 2013.

\bibitem{Ref16}
P.~Zhang, H.~Chen, L.~Zhang, T.~Ran, and D.~Zhang, ``Transparent self-cleaning
  lubricant-infused surfaces made with large-area breath figure patterns,''
  {\em Applied Surface Science}, vol.~355, pp.~1083--1090, 2015.

\bibitem{Ref17}
L.~Chen, A.~Geissler, E.~Bonaccurso, and K.~Zhang, ``Transparent slippery
  surfaces made with sustainable porous cellulose lauroyl ester films,'' {\em
  ACS Applied Materials \& Interfaces}, vol.~6, no.~9, pp.~6969--6976, 2014.

\bibitem{gas2017drop}
A.~Keiser, L.~Keiser, C.~Clanet, and D.~Qu{\'e}r{\'e}, ``Drop friction on
  liquid-infused materials,'' {\em Soft matter}, vol.~13, no.~39,
  pp.~6981--6987, 2017.

\bibitem{Ref18}
P.~Kim, T.-S. Wong, J.~Alvarenga, M.~J. Kreder, W.~E. Adorno-Martinez, and
  J.~Aizenberg, ``Liquid-infused nanostructured surfaces with extreme anti-ice
  and anti-frost performance,'' {\em ACS Nano}, vol.~6, no.~8, pp.~6569--6577,
  2012.

\bibitem{Ref19}
C.~Lee, H.~Kim, and Y.~Nam, ``Drop impact dynamics on oil-infused
  nanostructured surfaces,'' {\em Langmuir}, vol.~30, no.~28, pp.~8400--8407,
  2014.

\bibitem{kim2016droplet}
J.-H. Kim and J.~P. Rothstein, ``Droplet impact dynamics on lubricant-infused
  superhydrophobic surfaces: The role of viscosity ratio,'' {\em Langmuir},
  vol.~32, no.~40, pp.~10166--10176, 2016.

\bibitem{Ref20}
G.~Agbaglah, C.~Josserand, and S.~Zaleski, ``Longitudinal instability of a
  liquid rim,'' {\em Physics of Fluids}, vol.~25, no.~2, p.~022103, 2013.

\bibitem{Ref21}
F.~Guillemot, A.~Brunet-Bruneau, E.~Bourgeat-Lami, J.-P. Boilot, E.~Barthel,
  and T.~Gacoin, ``Percolation transition in the porous structure of
  latex-templated silica monoliths,'' {\em Microporous and Mesoporous
  Materials}, vol.~172, pp.~146--150, 2013.

\bibitem{Ref22}
J.~Zhang, L.~Wu, B.~Li, L.~Li, S.~Seeger, and A.~Wang, ``Evaporation-induced
  transition from nepenthes pitcher-inspired slippery surfaces to lotus
  leaf-inspired superoleophobic surfaces,'' {\em Langmuir}, vol.~30, no.~47,
  pp.~14292--14299, 2014.

\bibitem{Ref23}
P.~Kim, M.~J. Kreder, J.~Alvarenga, and J.~Aizenberg, ``Hierarchical or not?
  effect of the length scale and hierarchy of the surface roughness on
  omniphobicity of lubricant-infused substrates,'' {\em Nano Letters}, vol.~13,
  no.~4, pp.~1793--1799, 2013.

\bibitem{Ref24}
S.~Sunny, N.~Vogel, C.~Howell, T.~L. Vu, and J.~Aizenberg, ``Lubricant-infused
  nanoparticulate coatings assembled by layer-by-layer deposition,'' {\em
  Advanced Functional Materials}, vol.~24, no.~42, pp.~6658--6667, 2014.

\bibitem{Ref25}
D.~Bartolo, C.~Josserand, and D.~Bonn, ``Retraction dynamics of aqueous drops
  upon impact on non-wetting surfaces,'' {\em Journal of Fluid Mechanics},
  vol.~545, pp.~329--338, 2005.

\bibitem{Ref26}
C.~Clanet, C.~B\'eguin, D.~Richard, and D.~Qu\'er\'e, ``Maximal deformation of
  an impacting drop,'' {\em Journal of Fluid Mechanics}, vol.~517,
  pp.~199--208, 2004.

\bibitem{Ref27}
A.~L. Yarin, ``Drop impact dynamics: Splashing, spreading, receding,
  bouncing,'' {\em Annual Review of Fluid Mechanics}, vol.~38, no.~1,
  pp.~159--192, 2005.

\bibitem{Ref28}
C.~Josserand and S.~T. Thoroddsen, ``Drop impact on a solid surface,'' {\em
  Annual Review of Fluid Mechanics}, vol.~48, no.~1, pp.~365--391, 2016.

\bibitem{Ref29}
J.~C. Bird, R.~Dhiman, H.-M. Kwon, and K.~K. Varanasi, ``Reducing the contact
  time of a bouncing drop,'' {\em Nature}, vol.~503, no.~7476, pp.~385--388,
  2013.

\bibitem{Ref30}
X.~Li, X.~Ma, and Z.~Lan, ``Dynamic behavior of the water droplet impact on a
  textured hydrophobic/superhydrophobic surface: The effect of the remaining
  liquid film arising on the pillars' tops on the contact time,'' {\em
  Langmuir}, vol.~26, no.~7, pp.~4831--4838, 2010.

\bibitem{Ref31}
D.~Richard, C.~Clanet, and D.~Qu\'er\'e, ``Surface phenomena: Contact time of a
  bouncing drop,'' {\em Nature}, vol.~417, no.~6891, pp.~811--811, 2002.

\bibitem{antonini2013drop}
C.~Antonini, F.~Villa, I.~Bernagozzi, A.~Amirfazli, and M.~Marengo, ``Drop
  rebound after impact: the role of the receding contact angle,'' {\em
  Langmuir}, vol.~29, no.~52, pp.~16045--16050, 2013.

\bibitem{Ref32}
Y.~Shen, J.~Tao, H.~Tao, S.~Chen, L.~Pan, and T.~Wang, ``Relationship between
  wetting hysteresis and contact time of a bouncing droplet on hydrophobic
  surfaces,'' {\em ACS Applied Materials \& Interfaces}, vol.~7, no.~37,
  pp.~20972--20978, 2015.

\bibitem{malavasi2016knowledge}
I.~Malavasi, F.~Veronesi, A.~Caldarelli, M.~Zani, M.~Raimondo, and M.~Marengo,
  ``Is a knowledge of surface topology and contact angles enough to define the
  drop impact outcome?,'' {\em Langmuir}, vol.~32, no.~25, pp.~6255--6262,
  2016.

\end{thebibliography}
    \bibliographystyle{ieeetr}

    \end{document}